\documentclass[twocolumn,showpacs,amsmath,amssymb,prl]{revtex4}
\usepackage{graphicx}% Include figure files
\usepackage{dcolumn}% Align table columns on decimal point
\usepackage{bm}% bold math

\begin{document}
\input{epsf}

\title{21cm Absorption by Compact Hydrogen Disks Around \\ Black
Holes in Radio-Loud Nuclei of Galaxies}

\author{Abraham Loeb}

\affiliation{Astronomy Department, Harvard University, 60 Garden St.,
Cambridge, MA 02138, USA}

\begin{abstract} 

The clumpy maser disks observed in some galactic nuclei mark the outskirts
of the accretion disk that fuels the central black hole and provide a
potential site of nuclear star formation. Unfortunately, most of the gas in
maser disks is currently not being probed; large maser gains favor paths
that are characterized by a small velocity gradient and require rare
edge-on orientations of the disk. Here we propose a method for mapping the
atomic hydrogen distribution in nuclear disks through its 21cm absorption
against the radio continuum glow around the central black hole. In NGC
4258, the 21cm optical depth may approach unity for high angular-resolution
(VLBI) imaging of coherent clumps which are dominated by thermal broadening
and have the column density inferred from X-ray absorption data, $\sim
10^{23}~{\rm cm^{-2}}$.  Spreading the 21cm absorption over the full
rotation velocity width of the material in front of the narrow radio jets
gives a mean optical depth of $\sim 0.1$. Spectroscopic searches for the
21cm absorption feature in other galaxies can be used to identify the large
population of inclined gaseous disks which are not masing in our direction.
Follow-up imaging of 21cm silhouettes of accelerating clumps within these
disks can in turn be used to measure cosmological distances.

\end{abstract}

\pacs{98.54.Aj, 98.62.Js, 95.30.Jx, 98.80.-k}
\date{\today}
\maketitle

\section{I. Introduction}

The discovery of an edge-on H$_2$O maser disk in the nucleus of NGC 4258
\cite{Miyoshi,Moran} provides the best current evidence for a nuclear black
hole outside the Milky-Way galaxy. The radial extent of the maser ring,
$\sim 0.14$--0.28pc, and its measured Keplerian rotation profile between
$\sim 770$--$1100~{\rm km~s^{-1}}$, imply a central black hole mass 
$M\approx 4\times 10^7M_\odot$ \cite{He}.  To obtain large maser gains, any
such disk needs to be viewed at a nearly edge-on orientation with the maser
clumps lying along the lines of minimal variation in the line-of-sight
velocity \cite{Neu,Neu2}.  Given the small geometric likelihood of edge-on
orientations, the subsequent discovery of maser disks in the nuclei of
other galaxies \cite{Kond,Braa,Braa2}, demonstrated that cold nuclear disks
must be ubiquitous. The maser clumps potentially mark the outskirts of the
accretion disk that fuels the central black hole in these galaxies.
%Interestingly, the maser clumps also reside within the same logarithmic bin
%of radius where the broad emission lines of quasars are generated.

The Milky Way galaxy does not show evidence for a compact nuclear disk but
instead features young massive stars at an orbital radius of $\lesssim
0.3~{\rm pc}$ from the central black hole, SgrA* \cite{Ghez,Genzel}.  These
stars could not have formed in a central molecular cloud because of the
strong tidal field of SgrA* \cite{Morris}. Instead, they are conjectured to
have formed in a cold compact disk, at a sub-parsec distance from SgrA*,
where the disk self-gravity was important \cite{Levin,Genzel2,ML}. For the
generic radial profile of a viscous accretion disk, the Toomre
$Q$-parameter is indeed expected to approach unity at the required distance
scale \cite{Goodman,Tan,Sunyaev, Shlosman}, making the disk unstable for
fragmentation. This coincidence could also explain the clumpy nature of
maser disks at a rotation velocity of $\sim 10^3~{\rm km~s^{-1}}$
\cite{ML}.  Elsewhere in our Galaxy, masers are observed to be associated
with regions in which massive stars form \cite{Reid,Elitzur}.  It is
therefore likely that the observed maser disks in external galactic nuclei
represent sites of star formation \cite{ML}.  The abundance of maser disks
implies that galactic nuclei experience common episodes of disk formation,
during which massive stars -- like those observed near SgrA*, are born.
Winds from these massive stars or supernova explosions (supplemented by
additional feedback from the accreting nuclear black hole) could have
dispersed the central disk in the Milky-Way galaxy on a short timescale of
$\lesssim 10^7~{\rm years}$. In the future, a new disk might form when
fresh cold gas will assemble once again around SgrA*.

Little is known observationally about the global gas distribution in the
observed maser disks \cite{He}. Here we propose to map atomic hydrogen in
nuclear disks through its 21cm absorption of background radio glow around
the black hole.  It is known that radio-emitting plumes (which define the
base of large-scale jets \cite{Cec}) exist around NGC 4258 (see Fig. 1 in
Ref. \cite{He}), but the angular extent of the coronal radio emission at
$1.42$GHz and its potential absorption by the disk have not been explored
as of yet in the published literature \cite{Unpub}. In \S 2 we calculate
the expected optical depth for 21cm absorption in nuclear gaseous disks.
This absorption signature could be identified through Very Long Baseline
Interferometry (VLBI) observations of regions where the continuum
backlighting is sufficiently bright.

In principle, the 21cm absorption by nuclear disks can be mapped at high
angular and spectral resolutions. The velocity and acceleration of clumps
within the disk can then be used to infer the angular diameter distance to
the sources, as demonstrated for maser clumps in NGC 4258
\cite{Greenhill,Maoz,Moran,Hump}.  The 21cm absorption feature can also be
searched for spectroscopically (without spatial resolution) in a survey
over a large number of compact radio sources.  Because the 21cm absorption
signature would appear for arbitrary disk inclination, a dedicated search
for this signature would be more likely to find nuclear disks than searches
for masers which are limited to edge-on orientations of the
disks. Follow-up VLBI imaging of disks could then be used to infer the
central black hole mass for a large sample of galaxies.

\section{II. Optical Depth of compact nuclear disks}

The radiative transfer equation for the intensity $I_\nu$ of the 21cm line
along a particular line-of-sight reads \cite{RL},
\begin{equation}
{dI_\nu\over ds}={\phi(\nu) h\nu\over 4\pi}\left[n_2A_{21} -
\left(n_1B_{12} -n_2B_{21}\right)I_\nu\right],
\label{rad}
\end{equation}
where $\nu$ is the photon frequency, $ds$ is the path element, $\phi(\nu)$
is the line profile function normalized by $\int \phi(\nu)d\nu=1$ (with an
amplitude of order the inverse of the frequency width of the line, $\Delta
\nu$), subscripts 1 and 2 denote the lower and upper levels of the line,
$n$ denotes the number density of atoms at the different levels, and $A$
and $B$ are the Einstein coefficients for the transition between these
levels. We make use of the standard relations: $B_{21}=(g_1/g_2)B_{12}$ and
$B_{12}=(g_2/g_1)A_{21}n/I_\nu$, where $g$ is the spin degeneracy factor of
each state. For the 21cm transition, $A_{21}=2.85\times 10^{-15} {\rm
s^{-1}}$ and $g_2/g_1=3$ \cite{Field}.  The relative populations of
hydrogen atoms in the two spin states defines the so-called spin
temperature, $T_s$, through the relation, $(n_2/n_1)=
(g_2/g_1)\exp\{-E/kT_s\}$, where $E/k=0.068$K is the transition energy. In
the regime of interest here, $T_s\gg E$ and so $[(g_2/g_1) (n_1/n_2)
-1]\approx E/kT_s$, and $n_2\approx {3\over 4} n_H$, where $n_H$ is the
total number density of hydrogen atoms. Moreover, the brightness of the
spontaneous 21cm emission is too weak to be detectable. We therefore ignore
the first term in the square brackets of Eq. (\ref{rad}) and consider the
absorption signature of the gas against a bright radio continuum glow in
the background.  Defining the optical depth along a ray as $\tau=-{\Delta
\ln I_\nu}$, we get
\begin{equation}
\tau(\nu)={3\over 32\pi} {h^3c^2 A_{21}\over E^2}\left[\nu \phi(\nu)\right]
{N_H\over kT_s},
\label{tau}
\end{equation}
where $N_H=\int n_H ds$ is the column density of hydrogen.  

To get an estimate for the average optical depth value across the line
profile, we write $\nu\phi(\nu)=(\Delta \nu/\nu)^{-1}$, where $(\Delta
\nu/\nu_0)=(\Delta v/c)$, is the fractional Doppler width of the line,
corresponding to a velocity spread $\Delta v$ among the absorbing
atoms. The minimum line width attainable is dictated by the spread in the
thermal velocities of the atoms, for which \cite{RL}, $\Delta v=v_{\rm th}=
(2kT/m_H)^{1/2}$, where $m_H$ is the hydrogen atom mass. A larger width can
be induced by a gradient in the bulk velocity of the gas along the
line-of-sight and is calculated in the Sobolev approximation
\footnote{In the Sobolev approximation \cite{Sobolev}, the term
$\nu\phi(\nu)N_H$ in Eq. (\ref{tau}) is replaced by $c
n_H/(dv_{\parallel}/ds)$ where $dv_{\parallel}/ds$ is the line-of-sight
gradient of the line-of-sight velocity of the gas.  In the cosmological
context of 21cm absorption by a uniform intergalactic medium which
encounters Hubble expansion, this gradient is simply the Hubble parameter
(see, e.g., Ref. \cite{Bar}).}.  When the 21cm line is optically-thin
($\tau \lesssim 1$), the absorption signal obtains a width that reflects
all the contributing gas elements within the angular resolution and
frequency band of the observations.

Substituting all the coefficients into Eq. (\ref{tau}) yields,
\begin{equation}
\tau = 0.7 \left({N_H\over 10^{23}~{\rm cm^{-2}}}\right)\left({T_s\over
8\times 10^3~{\rm K}}\right)^{-1}\left({\Delta v\over 10~{\rm
km~s^{-1}}}\right)^{-1}.
\label{opt}
\end{equation}
The specific numerical values that were substituted on the right-hand-side
of Eq. (\ref{opt}), correspond to the expected parameters of the atomic
hydrogen disk in NGC 4258 \cite{He}.  X-ray observations of this system
indicate a hydrogen column density of $\sim 10^{23}~{\rm cm^{-2}}$
\cite{xray}.  Based on the observed X-ray luminosity of NGC 4258 and the
warped geometry of its disk, the gas is expected to be predominantly atomic
(rather than molecular) outside a radius $\sim 0.3~{\rm pc}$ (see Fig. 23
in Ref. \cite{He}). Theoretical calculations \cite{Neu2} suggest that the
atomic hydrogen \footnote{The molecular gas in the disk is cooler ($\sim
10^3$ K) and could potentially be probed through other absorption
lines.}  in the disk of NGC 4258 has an asymptotic temperature $\sim
8000~{\rm K}$, providing a thermal velocity width of $v_{\rm th}\approx
10~{\rm km~s^{-1}}$. At the high densities under consideration, we assume
that the spin temperature $T_s$ is in collisional equilibrium with the
kinetic temperature of the gas.  High resolution imaging of nuclear disks
can therefore be used to constrain their density and temperature
distributions through Eq. (\ref{opt}).

The scale height of a thin accretion disk, $h$, is expected \cite{ST} to be
a fraction $\sim (v_{\rm th}/v_\phi)$ of its radius $r$, where
$v_\phi(r)=(GM/r)^{1/2}$ is the rotation speed at $r$ for a black hole mass
$M$.  A line of sight which crosses the disk at an angle $\theta$ relative
to the normal to the disk samples a spread in the line-of-sight bulk
velocity that is $\leq (2h/\cos\theta)[(dv_\phi/dr)\sin\theta] =
(\sin^2\theta/\cos\theta) v_{\rm th}$. For the warped (bowl-shaped) maser
region of NGC 4258, the value of $\cos\theta\approx 0.3$ \cite{He} yields
$\Delta v\sim (1$--$3)\times v_{\rm th}$.

In order for VLBI imaging to reach the thermal broadening minimum of
$\Delta v$, a particular spatial resolution element needs to be dominated
by a single clump of gas with a coherent bulk velocity.  Such a clump would
have a small line-of-sight bulk velocity if it is located in front of the
black hole and up to the full rotation speed on the side.  If the clump
fills only a fraction $f$ of the source area within the resolution element,
then $\tau$ will be reduced by a factor of $f$.  For the diffuse gas in the
disk, the spatial resolution required to achieve the thermal width minimum
is of order the disk scale height, $h\sim (v_{\rm th}/v_\phi) r$. This
resolution scale corresponds to $\sim 10^{-2} r\sim 3\times 10^{-3}~{\rm
pc}$ for NGC 4258. At a wavelength of 21cm and a source distance of 7.2Mpc,
this resolution requires an unrealistic baseline of $\sim 5\times 10^5$km,
larger by a factor of 40 than the diameter of the Earth.  Thus, a
terrestrial VLBI will resolve the disk in NGC 4258 only around and outside
the maser region ($r\gtrsim 0.14~{\rm pc}$). Coincidentally, this is indeed
the region expected to be dominated by atomic hydrogen \cite{He}. An
analogous disk around a quasar black hole whose mass $M$ is larger by two
orders of magnitude than in NGC 4258, could in principle be resolved out to
a distance of $\sim 1$ Gpc.

We conclude that the high value of the optical depth in Eq. (\ref{opt})
applies to silhouettes of coherent clumps in which thermal broadening
dictates the velocity spread $\Delta v$.  Such clumps are expected to exist
outside the maser region, where the Toomre-$Q$ parameter is of order unity
or lower \cite{ML}. If individual clumps of atomic hydrogen are not
resolved or if the disk is smooth, then the optical depth would be diluted
over a broader velocity width. In general, the absorption depth at a given
frequency bin scales as the fractional (brightness-weighted) area of the
continuum source over which hydrogen atoms resonate with photons in the
observed frequency bin.

Under a uniform background illumination, the absorption spectrum of an
unresolved circular disk can be obtained through a sum over concentric
rings in the disk plane. We assume that the normal to the disk plane is
inclined at an angle $\theta$ relative to the line-of-sight. A single
optically-thin ring with a circular rotation velocity
$v_\phi(r)=(GM/r)^{1/2}$ and a radius $r$, gives a U-shaped spectral
profile in terms of $-1<\delta (\nu)<1$ \cite{RH},
\begin{equation}
\tau(\nu,r)={\tau_{\rm ring}(r)\over \pi (1-\delta^2)^{1/2}},
\label{spread} 
\end{equation}
where $\delta=[(\nu-\nu_0)/\nu_0]/[0.5\Delta v/c]$,
$\Delta v=2v_\phi(r) \sin \theta$, and 
\begin{eqnarray} 
\tau_{\rm ring} (r)& = & 0.44\times 10^{-2} \left({N_H (r)\over 10^{23}~{\rm
cm^{-2}}}\right)\left({T_s\over 8\times 10^3~{\rm
K}}\right)^{-1}\nonumber \\
& \times & \left({v_\phi(r) \sin \theta \over 800~{\rm
km~s^{-1}}}\right)^{-1}.
\label{tauring}
\end{eqnarray}
The total absorption feature of an unresolved disk can be obtained by
summing over all the rings in which atomic hydrogen resides, weighted by
the brightness distribution of the backlighting at $\nu_0=1.42$GHz.

For an arbitrary background illumination, the net deficit in the fractional
spectral intensity across the area $S$ of an unresolved optically-thin
source is given by,
\begin{equation}
{\Delta I_S\over I_S}(\nu) = - {\int_S I_\nu(x,y) \tau(\nu,x,y) dx dy \over
\int_S I_\nu (x,y) dx dy} ,
\label{total}
\end{equation}
where $(x,y)$ are the sky coordinates and the unabsorbed continuum source
can be assumed to have a smooth (typically power-law, $I_\nu\propto
\nu^{\alpha}$) spectrum across the absorption line profile in the
numerator. For a thin disk which is not perfectly edge-on, the exact
expression for the optical depth $\tau(\nu,x,y)$ in Eq. (\ref{tau}) can be
approximated as the thermally broadened value in Eq. (\ref{opt}) at the
Doppler shifted frequency $\nu_0[1-v_{\parallel}/c]$, where
$v_\parallel(x,y)$ is the line-of-sight component of the bulk velocity of
the gas.  Clearly, in order for the spectral deficit to be noticeable, a
dominant component of the radio emission needs to originate behind the
absorbing disk. If the background illumination originates from a narrowly
collimated jet (as indicated by the 22GHz image of NGC 4258), then the
absorption feature will be characterized by the low line-of-sight velocity
spread ($\Delta v$) of the material in front of the jet. In this case, the
spectral deficit will be larger than the deficit associated with the full
velocity spread of the disk.  For the narrow jets of NGC 4258 \cite{He}, we
estimate $\Delta v\sim 100~{\rm km~s^{-1}}$ and $\tau\sim 0.1$. A dedicated
search for the 21cm spectroscopic feature in other compact radio sources
can be used to identify new nuclear disks in distant galaxies.

\section{III. Discussion}

The parameters of the maser disk in NGC 4258 imply that atomic hydrogen
within the compact gaseous disks in galactic nuclei can produce measurable
21cm absorption against the backlight of continuum radio emission around
the central black hole.  For NGC 4258, the 21cm optical depth may approach
unity in silhouettes of coherent clumps which are dominated by thermal
broadening and have the column density inferred from X-ray absorption data
\cite{xray}. Spreading the absorption across the rotation velocity width of
the material in front of the collimated jets in NGC 4258 results in an
expected optical depth of $\sim 0.1$.

More than half of all nuclear H$_2$O megamasers show X-ray absorption with
column densities $N_H \sim 10^{24}$--$10^{25}~{\rm cm}^{-2}$ (see
Ref. \cite{Mad} and Fig. 7 in Ref.  \cite{Zhang}), at which the optical
depth for 21cm absorption might exceed unity.  High-resolution VLBI images
of 21cm absorption can be used to map the distributions of the density,
temperature, and line-of-sight velocity of atomic hydrogen in nuclear
disks. Such maps could show direct evidence for spiral arms \cite{MM},
which are conjectured to exist based on the latest maser data in NGC
4258\cite{Hump,Moran}. More generally, the maps hold the potential for
testing current models of accretion disks, shedding light on the geometry
of obscured (Compton-thick) quasars, and improving our understanding of
star formation in galactic nuclei.

A comprehensive search for a spectroscopic absorption feature in radio-loud
nuclei of galaxies can be used to find a large number of inclined gaseous
disks which are not masing in our direction. The improved statistics of
known nuclear disks would provide better understanding of the duty cycle of
black hole fueling and star formation in galactic nuclei.

Remote atomic hydrogen within the host galaxy would also result in
absorption but will be limited to low-velocity widths. The compact nuclear
disk is expected to dominate the wings of the 21cm absorption profile which
extend out to velocity offsets of $\pm 10^3~{\rm km~s^{-1}}$. Follow-up
imaging of the nuclear disk can be used to separate out extended galactic
absorption.

VLBI measurements of the velocity and acceleration of coherent hydrogen
clumps within the disk can be used to infer the angular diameter distance
to the source, as demonstrated with maser clumps 
\cite{Greenhill,Maoz,Moran,Hump}. Detection of suitable radio sources out
to sizeable redshifts could potentially place new constraints on the
equation of state of the dark energy through the dependence of the angular
diameter distance on source redshift \cite{Greenhill}. The selection of a
large number of suitable targets would be a particularly attractive goal
\cite{Brigg} for the planned Square Kilometer Array
\footnote{http://www.skatelescope.org/}.

\bigskip
\bigskip
\paragraph*{Acknowledgments.}
I thank Jim Moran and George Rybicki for useful discussions.

\end{document}